\documentclass[12pt]{article}
\usepackage[style=chem-acs,articletitle=true]{biblatex}
\usepackage[parfill]{parskip}

\usepackage{geometry}
\usepackage{float}
\usepackage{siunitx}
\usepackage{upgreek}
\usepackage{setspace}
\usepackage{etoc}
 \usepackage[font=small,font=sf, labelfont=bf]{caption}

\newcommand{\beginsupplement}{%
        \setcounter{section}{0}
        \renewcommand{\thesubsection}{\arabic{subsection}}%
        \setcounter{table}{0}
        \renewcommand{\thetable}{S.\arabic{table}}%
        \setcounter{figure}{0}
        \renewcommand{\thefigure}{S\arabic{figure}}%
        \setcounter{equation}{0}
        \renewcommand{\theequation}{S\arabic{equation}}%
     }

\usepackage[utf8]{inputenc}
\usepackage{mathtools}
\usepackage{amsmath}
\usepackage{braket}
\usepackage{textcomp}
\usepackage{xcolor}
\usepackage{hyperref}
\usepackage{bm}
\def\be{\begin{equation}}
\def\ee{\end{equation}}
\title{\Large{\textbf{Electrical spectroscopy of the spin-wave dispersion and bistability in gallium-doped yttrium iron garnet}}}
\author{{\normalsize Joris J. Carmiggelt}$^{1}$, {\normalsize Olaf C. Dreijer}$^{1}$, {\normalsize Carsten Dubs}$^{2}$, {\normalsize Oleksii Surzhenko}$^{2}$, {\normalsize Toeno van der Sar}$^{1,*}$}
\date{
\begin{flushleft}
\footnotesize{
$^1$Department of Quantum Nanoscience, Kavli Institute of Nanoscience, Delft University of Technology, 2628 CJ Delft, The Netherlands\\
$^2$INNOVENT e.V. Technologieentwicklung, D-07745 Jena, Germany\\
\hfill\break
$^*$ Corresponding author. Email: t.vandersar@tudelft.nl }
\end{flushleft}
}
\begin{document}
\maketitle
\begin{abstract}
Yttrium iron garnet (YIG) is a magnetic insulator with record-low damping, allowing spin-wave transport over macroscopic distances. Doping YIG with gallium ions greatly reduces the demagnetizing field and introduces a perpendicular magnetic anisotropy, which leads to an isotropic spin-wave dispersion that facilitates spin-wave optics and spin-wave steering. Here, we characterize the dispersion of a gallium-doped YIG (Ga:YIG) thin film using electrical spectroscopy. We determine the magnetic anisotropy parameters from the ferromagnetic resonance frequency and use propagating spin wave spectroscopy in the Damon-Eshbach configuration to detect the small spin-wave magnetic fields of this ultrathin weak magnet over a wide range of wavevectors, enabling the extraction of the exchange constant $\alpha=1.3(2)\times10^{-12}$ J/m. The frequencies of the spin waves shift with increasing drive power, which eventually leads to the foldover of the spin-wave modes. Our results shed light on isotropic spin-wave transport in Ga:YIG and highlight the potential of electrical spectroscopy to map out the dispersion and bistability of propagating spin waves in magnets with a low saturation magnetization. 
\end{abstract}
\newpage
\begin{refsection}
\addcontentsline{toc}{section}{Main Text}
Yttrium iron garnet (YIG) is a magnetic insulator that is famous for its low Gilbert damping and long-range spin-wave propagation~\cite{Chumak2010}. At low bias fields the YIG magnetization is typically pushed in the plane by the demagnetizing field~\cite{Guslienko2011}, leading to an anisotropic spin-wave dispersion at microwave frequencies. For applications that rely on spin-wave optics and spin-wave steering an isotropic spin-wave dispersion is desirable~\cite{Pirro2021}, which can be achieved by introducing gallium dopants in the YIG: The presence of the dopants reduces the saturation magnetization and thereby the demagnetizing field~\cite{Hansen1974}, and induces a perpendicular magnetic anisotropy (PMA)~\cite{Mee1971,Heinz1971}, such that the magnetization points out-of-plane. Isotropic transport of forward-volume spin waves has been observed even at zero bias field~\cite{Haldar2017}, opening the door for spin-wave logic devices~\cite{Ustinov2013,Klingler2015,Kanazawa2016}.\\

To harness isotropic spin waves it is essential to know the spin-wave dispersion, which is dominated by the exchange interaction for magnets with a low saturation magnetization~\cite{Kalinikos1986}. Here, we use all-electrical spectroscopy of propagating spin waves to characterize the spin-wave dispersion of a 45-nm-thick film of gallium-doped YIG (Ga:YIG). Rather than looking at the discrete mode numbers of perpendicular standing spin waves~\cite{Klingler2015a}, this method enables extracting the exchange constant by monitoring the spin-wave transmission for a continuously-tunable range of wavevectors. We show that this technique has sufficient sensitivity to characterize spin waves in nanometer-thick Ga:YIG films, where perpendicular modes may be challenging to detect due to their high frequencies and small mode overlap with the stripline drive field.\\

We extract the anisotropy parameters from the field dependence of the ferromagnetic resonance (FMR) frequency at different bias field orientations and find that the PMA is strong enough to lift the magnetization out of the plane. Next, we characterize the spin-wave dispersion from electrically-detected spin-wave spectra. We measure in the Damon-Eshbach configuration to boost the inductive coupling of the spin waves to the striplines~\cite{Bhaskar2020}, allowing the extraction of the spin-wave group velocity over a wide range of wavevectors from which we determine the exchange constant. When increasing the microwave excitation power, we observe clear frequency shifts of the spin-wave modes. The shifts result in the foldover of spin waves, which we verify by comparing upward and downward frequency sweeps. These results benchmark propagating spin wave spectroscopy as an accessible tool to characterize the exchange constant and spin-wave foldover in technologically attractive thin-film magnetic insulators with low saturation magnetization and PMA.\\

\begin{figure}[h!]\centering
	\includegraphics[scale=1.3]{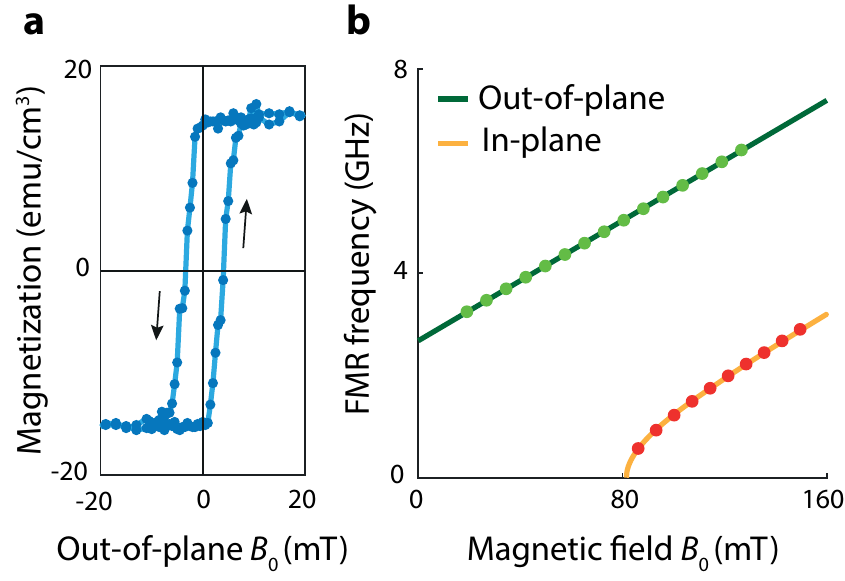}
	\caption{\textbf{The saturation magnetization and magnetic anisotropies of Ga:YIG.} (a) Hysteresis loop of the magnetization of a 45-nm-thick Ga:YIG film as a function of out-of-plane magnetic field $B_0$ measured using vibrating sample magnetometry and corrected for magnetic background. The arrows denote the sweep direction of the magnetic field. (b) FMR measurements using an out-of-plane (green) and in-plane (red) magnetic field $B_0$. From the fits of the FMR frequencies (solid lines) we determine the perpendicular and cubic anisotropy fields (see text).                 
	}
	\label{fig1}
\end{figure}

We use liquid phase epitaxy to grow a 45-nm-thick film of Ga:YIG on an (111)-oriented gadolinium gallium garnet (GGG) substrate (supplementary material section~\ref{Sup:GaYIG}). Using vibrating sample magnetometry (VSM) we determine the saturation magnetization $M_\text s=1.52(6)\times10^4$ A/m (Fig.~\ref{fig1}a, the number in parentheses denotes the 95\% confidence interval), which is approximately an order of magnitude smaller than undoped YIG films of similar thicknesses~\cite{Dubs2020}. \\

In addition to PMA, Ga:YIG films also have a cubic magnetic anisotropy due to a cubic unit cell. We start by determining the cubic and perpendicular anisotropy fields from the ferromagnetic resonance (FMR) frequencies $\omega_\text{FMR}/2\pi$ using an out-of-plane ($\perp$) and in-plane ($||$) magnetic bias field $B_0$. For (111)-oriented films the out-of-plane and in-plane Kittel relations are given by~\cite{Dubs2020,Kalinikos1990}
\be
\omega_\text{FMR($\perp$)}=\gamma_\perp(B_0-\mu_0M_\text s+\frac{2K_{2\perp}}{M_\text s}-\frac{4}{3}\frac{K_4}{M_\text s}),
\label{eq1}
\ee
\be
\omega_\text{FMR($||$)}=\gamma_{||}\sqrt{B_0(B_0+\mu_0M_\text s-\frac{2K_{2\perp}}{M_\text s}-\frac{K_4}{M_\text s})}.
\label{eq2}
\ee
Here $\gamma_{\perp,||}=g_{\perp,||}\mu_\text B/\hbar$ is the gyromagnetic ratio with $g_{\perp,||}$ the anisotropic g-factor, $\mu_\text B$ the Bohr magneton and $\hbar$ the reduced Planck constant, $\mu_0$ is the magnetic permeability of free space, $K_{2\perp}$ is the uniaxial out-of-plane anisotropy (e.g. PMA) constant and $K_4$ the cubic anisotropy constant. During the in-plane FMR measurement we apply the magnetic field along the $[1\overline{1}0]$ crystallographic axis to minimize the out-of-plane component of the magnetization (supplementary material section~\ref{Sup:FMR}). We neglect any uniaxial in-plane anisotropy as it is known to be small in YIG samples~\cite{Dubs2020}.\\

By substituting the value of $M_\text s$ that we obtained with VSM into equations~\ref{eq1} and~\ref{eq2}, we can determine $K_{2\perp}$ and $K_4$ from the FMR frequencies (Fig.~\ref{fig1}b)~\cite{Manuilov2009}. From the fits (solid lines) we extract the uniaxial out-of-plane anisotropy field $2K_{2\perp}/M_\text s=104.7(8)$ mT and the cubic anisotropy field $2K_{4}/M_\text s=-8.2(5)$ mT (supplementary material section~\ref{Sup:meas}). Undoped YIG films of similar thicknesses have comparable cubic anisotropy fields~\cite{Dubs2020}, which agrees with previous work on micrometer-scale films showing that the cubic anisotropy of YIG does not depend on gallium concentration~\cite{Bobkov2003}. We determine the in-plane and out-of-plane g-factors to be $g_{||}=2.041(4)$ and $g_\perp=2.101(3)$~\cite{Farle1998}.\\

\begin{figure}[h!]\centering
	\includegraphics[scale=1.3]{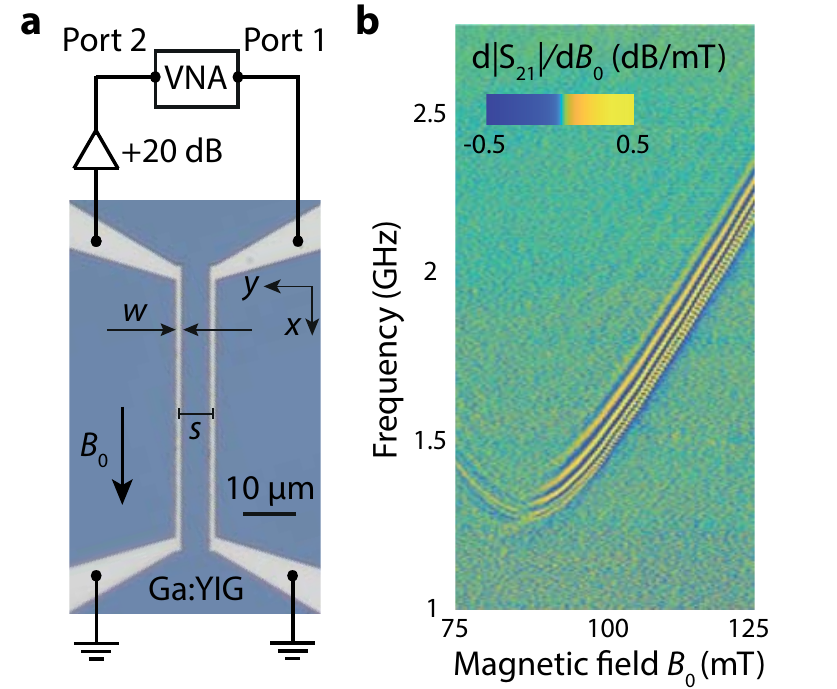}
	\caption{\textbf{All-electrical propagating spin wave spectroscopy.} (a) Optical micrograph of a Ga:YIG film with two gold striplines that are connected to the ports of a vector network analyser (VNA). Port 1 applies a microwave current (typical excitation power: $-35$ dBm) that induces a radio-frequency magnetic field $B^\text{RF}$ at the injector stripline. This field excites propagating spin waves that couple inductively to the detector stripline at a distance $s$. The generated microwave current is amplified and detected at port 2. A static magnetic field $B_0$ is applied in the Damon-Eshbach configuration and is oriented such that the chirality of $B^\text{RF}$ favours the excitation of spin waves propagating towards the detector stripline~\cite{Yu2019}. (b) Field-derivative of the microwave transmission $|\text S_\text{21}|$ between two striplines ($w=1$ $\upmu$m, $s=6$ $\upmu$m) as a function of $B_0$ and microwave frequency. The colormap is squeezed, such that also fringes corresponding to low-amplitude spin waves are visible. A masked background was subtracted to highlight the signal attributed to spin waves (supplementary material section~\ref{Sup:BG}).                 
	}
	\label{fig2}
\end{figure}

We now use propagating spin wave spectroscopy to characterize the spin-wave dispersion in Ga:YIG. We measure the microwave transmission $|\text S_\text{21}|$ between two microstrips fabricated directly on the Ga:YIG as a function of static magnetic field $B_0$ and frequency $f$ (Fig.~\ref{fig2}a). The magnetic field is applied in the Damon-Eshbach geometry to maximize the inductive coupling between the spin waves and the striplines~\cite{Bhaskar2020}. We measure a clear Damon-Eshbach spin-wave signal in the microwave transmission spectrum when $B_0$ overcomes the PMA and pushes the spins in the plane (Fig.~\ref{fig2}b, supplementary material section~\ref{Sup:BG}). The signal appears at a finite frequency, because the bias field $B_0$ is applied along the $[11\overline{2}]$ crystallographic axis with a finite out-of-plane angle of $\sim1^\circ$ (supplementary material section~\ref{Sup:FMR}).\\

\begin{figure}[h!]\centering
	\includegraphics[scale=1.3]{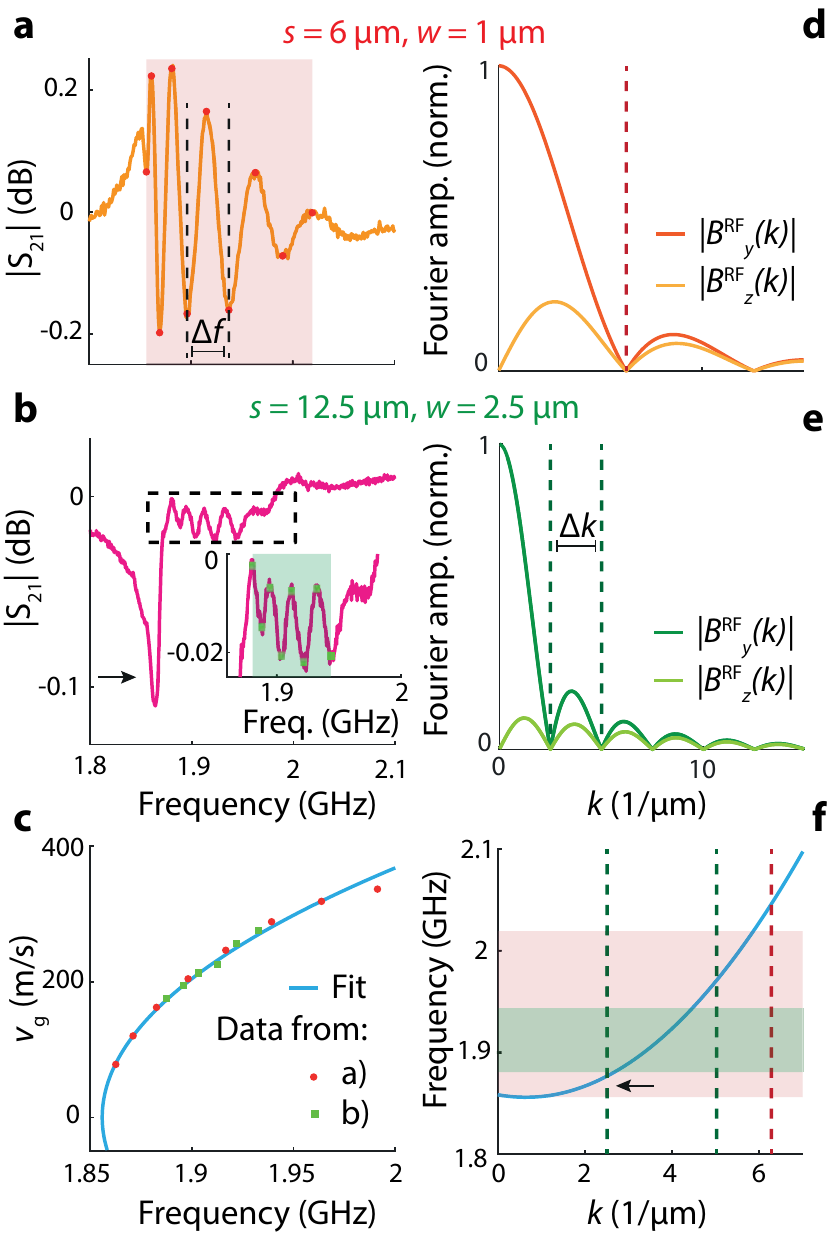}
	\caption{\textbf{Extracting the exchange constant from spin-wave transmission spectra.} (a,b) Background-subtracted linetraces of $|\text S_\text{21}|$ for two sets of striplines (a: $w=1$ $\upmu$m, $s=6$ $\upmu$m, b: $w=2.5$ $\upmu$m, $s=12.5$ $\upmu$m, excitation power: $-35$ dBm). The red circles (a) and green squares (inset of b) mark the extrema of the spin-wave fringes. (c) From the frequency difference between the extrema $\Delta f$ we determine the group velocity $v_\text g$ of the spin waves at the center frequency between the extrema. The blue line fits the data with an analytical expression for $v_\text g$, extracting the exchange constant $\alpha=1.3(2)\times10^{-12}$ J/m. (d,e) Normalized Fourier amplitude of the $y$ and $z$ components of the microwave excitation field $B^\text{RF}$ for striplines with widths $w=1$ $\upmu$m (d) and $w=2.5$ $\upmu$m (e). (f) Reconstructed spin-wave dispersion based on the fit in (c). The shaded areas correspond to the frequencies of the extrema in (a,b). The dashed lines are the same as in (d,e) and indicate the nodes in $|B^\text{RF}(k)|$ of the striplines. Only spin waves that are efficiently excited and detected by the striplines are observed in (a,b).                 
	}
	\label{fig3}
\end{figure}

The fringes in the transmission spectra result from the interference between the spin waves and the microwave excitation field~\cite{Ciubotaru2016,Bertelli2020}. Each fringe indicates an extra spin-wavelength $\lambda$ that fits between the striplines. We can thus use the fringes to determine the group velocity $v_\text g$  of the spin waves via~\cite{Neusser2010}
\be
v_\text g=\frac{\partial\omega_\text{SW}}{\partial k}\approx\frac{2\pi\Delta f}{2\pi/s}=\Delta fs.
\label{eq3}
\ee
Here $\omega_\text{SW}=2\pi f$ and $k=2\pi/\lambda$ are the spin wave’s angular frequency and wavevector, $\Delta f$ is the frequency difference between two consecutive maxima or minima of the fringes (Fig.~\ref{fig3}a) and $s$ is the center-to-center distance between both microstrips.\\

We extract the exchange constant of our Ga:YIG film by fitting the measured group velocity to an analytical expression derived from the spin-wave dispersion. The Damon-Eshbach spin-wave dispersion for magnetic thin films with cubic and perpendicular anisotropy is given by~\cite{Kalinikos1990} (supplementary material section~\ref{Sup:disp}) 
\be
\omega_\text{SW}(k)=\sqrt{\omega_B(\omega_B+\omega_M-\omega_K)+\frac{\omega_Mt}{2}(\omega_M-\omega_K)k+\gamma_{||}D(2\omega_B+\omega_M-\omega_K)k^2+\gamma_{||}^2D^2k^4}.
\label{eq4}
\ee
Here we defined for notational convenience $\omega_B=\gamma_{||}B_\text{0}$, $\omega_M=\gamma_{||}\mu_0M_\text s$, $\omega_{D}=\frac{\gamma_{||} D}{M_\text s}$, and $\omega_K =\gamma_{||}( 2K_{2\perp}/M_\text s+K_{4}/M_\text s)$, $t$ is the thickness of the film and $D=2\alpha/M_\text s$ is the spin stiffness, with $\alpha$ the exchange constant. Differentiating with respect to $k$ gives an analytical expression for the group velocity
\be
v_\text g(k)=\frac{1}{2\sqrt{\omega_\text{SW}(k)}}\big(\frac{\omega_Mt}{2}(\omega_M-\omega_K)+2\gamma_{||}D(2\omega_B+\omega_M-\omega_K)k+4\gamma_{||}^2D^2k^3\big).
\label{eq5}
\ee
Since we determined $M_\text s$ and the anisotropy constants from the VSM and FMR measurements, the exchange constant is the only unknown variable in the dispersion. We determine the exchange constant from spin-wave spectra measured using two sets of striplines with different widths and line-to-line distances ($w=1$ $\upmu$m, $s=6$ $\upmu$m and $w=2.5$ $\upmu$m, $s=12.5$ $\upmu$m) at the same static field (Fig.~\ref{fig3}a,b). First we extract $v_\text g$ as a function of frequency from the extrema in the spin-wave spectra using equation~\ref{eq3} (Fig.~\ref{fig3}c). By then fitting the measured $v_\text g(f)$ using equations~\ref{eq4} and~\ref{eq5} (solid line in Fig.~\ref{fig3}c), we find $\alpha=1.3(2)\times10^{-12}$ J/m and $B_0=117.5(3)$ mT (supplementary material section~\ref{Sup:meas}). The determined exchange constant is about 3 times smaller compared to undoped YIG~\cite{Klingler2015a}, which is in line with earlier observations of a decreasing exchange constant with increasing gallium concentration in micrometer-thick YIG films~\cite{Boyle1997}. Simultaneously the spin stiffness is increased by about 3 times compared to undoped YIG~\cite{Klingler2015a} due to the strong reduction of the saturation magnetization. For large wavelengths the group velocity is negative as a result of the PMA in the sample.\\

The spin-wave excitation and detection efficiency depends on the absolute value of the Fourier amplitude of the radio-frequency magnetic field $B^\text{RF}$ generated by a stripline, which oscillates in $k$ with a period given by $\Delta k=2\pi/w$ (Fig.~\ref{fig3}e)~\cite{Ciubotaru2016,Bertelli2020}. To verify that the spin waves we observe are efficiently excited and detected by our striplines, we substitute the extracted exchange constant into equation~\ref{eq4} and plot the spin-wave dispersion (Fig.~\ref{fig3}f). The shaded areas correspond to the frequencies of the spin-wave fringes (Fig.~\ref{fig3}a,b) and the dashed lines indicate the nodes in $|B^\text{RF}(k)|$ of both striplines (Fig.~\ref{fig3}d,e). We conclude that the fringes in Fig.~\ref{fig3}a correspond to spin waves excited by the first maximum of $|B^\text{RF}(k)|$ and that the fringes in Fig.~\ref{fig3}b correspond to spin waves excited by the second maximum.\\

Surprisingly, we do not observe fringes in Fig.~\ref{fig3}b corresponding to the first maximum of $|B^\text{RF}(k)|$, but rather see a dip in this frequency range (arrows in Fig.~\ref{fig3}b,f). This can be understood by noting that the average frequency difference between the fringes would be smaller than the spin-wave linewidth (supplementary material section~\ref{Sup:line}). Low-amplitude fringes corresponding to small-wavelength spin waves excited by the second k-space maximum of the 1-$\upmu$m-wide stripline are also visible (Fig.~\ref{fig2}b, supplementary material section~\ref{Sup:fring}). These results demonstrate that the spin-wave dispersion in weak magnets can be reliably extracted using propagating spin wave spectroscopy by combining measurements on striplines with different widths and spacings.  \\ 

When strongly driven to large amplitudes, the FMR behaves like a Duffing oscillator with a bistable response. Such bistability could potentially be harnessed for microwave switching~\cite{Fetisov1999}. Foldover of the FMR and standing spin-wave modes has been studied for several decades~\cite{Fetisov1999,Gui2009,Li2019a}, but foldover of propagating spin waves was only observed before in active feedback rings~\cite{Janantha2017}, spin-pumped systems~\cite{Ando2012} and magnonic ring resonators~\cite{Wang2020}. We show that we can characterize the foldover of propagating spin waves in Ga:YIG thin films using our spectroscopy technique.\\

\begin{figure}[h!]\centering
	\includegraphics[scale=1.3]{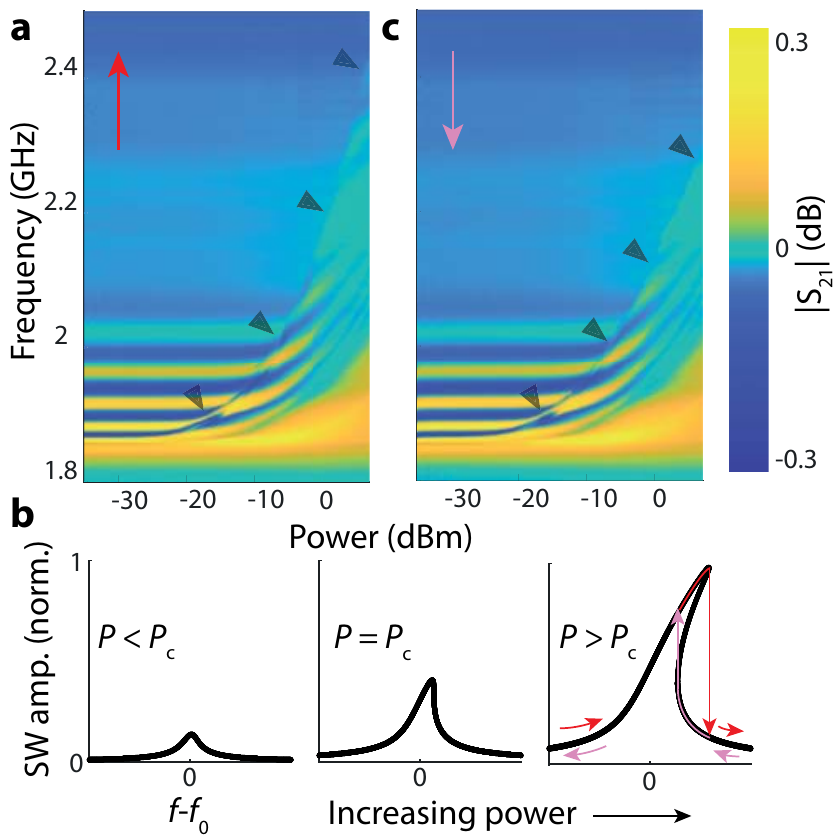}
	\caption{\textbf{Observation of spin-wave frequency shifts and foldover.} (a) Spin-wave spectra at different excitation powers ($w=1$ $\upmu$m, $s=6$ $\upmu$m). Low-frequency spin waves shift to higher frequencies when the microwave excitation power is increased. The markers indicate the sharp transition at which the spin waves fall back to their unshifted frequencies and serve as a guide to the eye. (b) Sketch of the normalized spin-wave amplitude vs frequency for increasing drive power $P$, showing the upward frequency shift away from the low-power resonance frequency $f_0$. Above a critical power $P_\text c$ the frequency shift results in the foldover of the spin-wave mode. As a result, the spin waves fall back to their unshifted dispersion at higher frequencies for upward frequency sweeps (red arrows, a) than downward sweeps (pink arrows, c).                  
	}
	\label{fig4}
\end{figure}

When increasing the drive power we observe frequency shifts of the spin waves (Fig.~\ref{fig4}a,c). These non-linear shifts result from the four-magnon self-interaction term in the spin-wave Hamiltonian. For an in-plane magnetized thin film, the shifts are given by~\cite{Krivosik2010}
\be
\tilde{\omega}_k=\omega_k+W_{kk,kk}|a_k|^2.
\ee
Here $\tilde{\omega}_k$ ($\omega_k$) is the non-linear (linear) spin-wave angular frequency, $W_{kk,kk}$ is the four-wave frequency-shift parameter and $a_k$ is the spin-wave amplitude. In our case $W_{kk,kk}$ is positive as a result of the PMA in the sample, leading to positive frequency shifts of the spin-wave modes (supplementary material section~\ref{Sup:freq}). The low-frequency spin waves start shifting first, because the stripline is the most efficient in exciting spin waves with small wavenumbers (Fig.~\ref{fig3}d,e). The spin-wave modes start shifting at a surprisingly low drive power of $\sim-30$ dBm, potentially caused by reduced spin-wave scattering~\cite{Li2019a} due to the low density of states associated with the increased spin stiffness and reduced saturation magnetization of our sample. \\

In the high-power microwave spectra we observe an abrupt transition at which the spin waves fall back to their unshifted low-power frequencies, indicating the foldover of the spin waves. As the spin-wave amplitude increases the spin-wave modes shift to higher frequencies, until the maximum amplitude is reached and the spin waves fall back to their low-amplitude dispersion (Fig.~\ref{fig4}b).\\

To demonstrate the foldover behaviour, we compare upward and downward frequency sweeps (Fig.~\ref{fig4}a,c). As expected the spin waves fall back to their unshifted dispersion earlier when sweeping against the frequency shift direction than when the sweep is in the same direction. The spin-wave amplitude and wavevector is thus bistable for the frequencies at which the foldover occurs. For these frequencies the stripline can excite two different wavelengths of spin waves at the same excitation power depending on the sweep direction that was used in the past.\\

The observed frequency shifts provide an extra knob for tuning the dispersion of spin waves. They give rise to strong non-linear microwave transmission between the striplines as a function of excitation power, which may provide opportunities for neuromorphic computing devices that simulate the spiking of artificial neurons above a certain input threshold~\cite{Wang2020,Feldmann2019}. \\

In summary, we used propagating spin wave spectroscopy to characterize the spin-wave dispersion in a 45-nm-thick film of Ga:YIG. The gallium doping reduces the saturation magnetization of the YIG and introduces a small PMA that lifts the magnetization out of the plane and causes the dispersion to be dominated by the exchange constant. We extract the exchange constant by fitting the group velocity at different frequencies and demonstrate that the detected spin waves are efficiently excited by the excitation fields of the striplines. Finally, we observe pronounced power-dependent frequency shifts of the spin waves that lead to foldover and mode bistability. Our results highlight the potential of all-electrical spectroscopy to shed light on the dispersion and nonlinear response of propagating spin waves in weakly-magnetic thin films.

\textbf{Supplementary material:} See the supplementary material for methods, details on the data analysis and error estimations, additional measurements and calculations of the FMR frequency, spin-wave dispersion and non-linear frequency-shift parameter.\\
\textbf{Author contributions:} J.J.C. and T.v.d.S. conceived the experiment. J.J.C. and O.D. built the experimental setup, performed the experiments and analyzed the data. C.D. grew the Ga:YIG film and O.S. performed the VSM measurement. J.J.C. fabricated the striplines. J.J.C. and T.v.d.S. wrote the manuscript with contributions from all coauthors. T.v.d.S. supervised the project. \\
\textbf{Acknowledgements:} This work was supported by the Netherlands Organisation for Scientific Research (NWO/OCW), as part of the Frontiers of Nanoscience program and by the Deutsche Forschungsgemeinschaft (DFG, German Research Foundation) -271741898. The authors thank A.V. Chumak for reviewing the manuscript, A. Katan, E. Lesne for useful discussions and C.C. Pothoven for fabricating the magnet holders used in the experimental setup. We also thank the staff of the TU Delft electronic support division and the Kavli Nanolab Delft for their support in soldering the printed circuit board and fabricating the microwave striplines.\\
\textbf{Competing interests:} The authors declare that they have no competing interests.\\
\textbf{Data availability:} All data contained in the figures are available in Zenodo.org at \url{http://doi.org/10.5281/zenodo.5494466}, reference number~\cite{Carmiggelt2021}. Additional data related to this paper are available from the corresponding author upon reasonable request.
\phantomsection\addcontentsline{toc}{subsection}{References}
\printbibliography
\end{refsection}
\newpage

\beginsupplement 
\setstretch{1.75}
\begin{refsection} 
\begin{center}
\textbf{{\huge Supplementary material}}\break\break
\textbf{{\Large\textbf{ Electrical spectroscopy of the spin-wave dispersion and bistability in gallium-doped yttrium iron garnet}}}\break\break
{\normalsize Joris J. Carmiggelt}$^{1}$, {\normalsize Olaf C. Dreijer}$^{1}$, {\normalsize Carsten Dubs}$^{2}$, {\normalsize Oleksii Surzhenko}$^{2}$,{\normalsize Toeno van der Sar}$^{1,*}$
\end{center}
\begin{flushleft}
\footnotesize{
$^1$Department of Quantum Nanoscience, Kavli Institute of Nanoscience, Delft University of Technology, 2628 CJ Delft, The Netherlands\\
$^2$INNOVENT e.V. Technologieentwicklung, D-07745 Jena, Germany\\
$^*$ Corresponding author. Email: t.vandersar@tudelft.nl }
\end{flushleft}
\phantomsection
\addcontentsline{toc}{section}{Supplementary material}
\localtableofcontents
\subsection{Ga:YIG sample and experimental setup}\label{Sup:GaYIG}
A 45-nm-thick film of gallium-doped yttrium iron garnet (Ga:YIG) was grown using liquid phase epitaxy on a one-inch (111) gadolinium gallium garnet (GGG) substrate and cut into chips of 5x5x0.5 mm$^3$. Striplines were fabricated on top of the Ga:YIG by e-beam lithography using PMMA(A8 495)/PMMA(A3 950) bilayer resist with an Elektra92 coating to avoid charging, and subsequent evaporation of Ti/Au (10 nm/190 nm). We wirebond the striplines to a printed circuit board and connect them to our vector network analyser (VNA, Keysight, P9372A) via small, non-magnetic SMPM connectors (Amphenol RF, 925-169J-51PT) to minimize spurious magnetic-field dependent signals and maximize the dynamic range of the bias field. Before reaching the VNA, the signals are amplified by a low noise +20 dB amplifier (Minicircuits, ZX60-83LN-S+) to avoid detection noise on the order of our signals. We place the sample between two large cylindrical permanent magnets (Supermagnete, S35-20-N) to apply a strong and homogeneous bias field. The magnets sit in home-built magnet holders that are mounted on computer-controlled translation stages (Thorlabs, MTS25-Z8, 25 mm range), which allow sweeping the field. We calibrate the magnetic field using a Hall probe (Hirst Magnetic Instruments, GM08). All measurements were performed at room temperature.

\subsection{Effect of the magnetic field alignment on the FMR frequency}\label{Sup:FMR}
\begin{figure}[h!]\centering
	\includegraphics[width=1\textwidth]{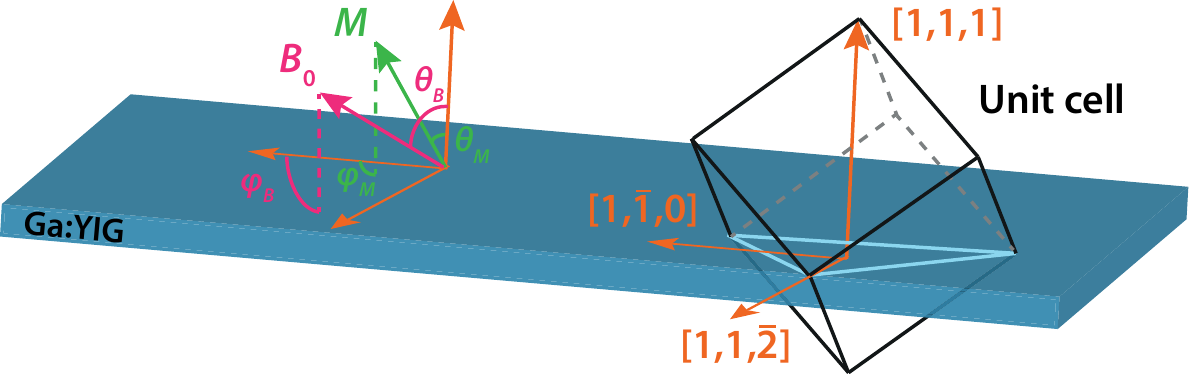}
	\caption{\textbf{Coordinate frame and crystallographic axes in Ga:YIG.} The $[1\overline{1}0]$ axis is slightly displaced to highlight the triangular symmetry plane (light blue) of the $(111)$-oriented cubic unit cell. After \cite{Manuilov2009}.                  
	}
	\label{figaxis}
\end{figure}

In this section we show that for (111)-oriented lattices with cubic anisotropy the in-plane Kittel relation holds when a strong magnetic field $B_0$ is applied along the in-plane $[1\overline{1}0]$ crystallographic axis. We also investigate the effect of a $\sim1^\circ$ out-of-plane angle of $B_0$ on the FMR frequency and the magnetization direction. Such a small angle may be present due to the manual placement of the sample in our setup (section~\ref{Sup:GaYIG}). \\
The FMR frequency is calculated according to~\cite{Suhl1955}
\be
\omega_\text{FMR}^2=\frac{\gamma^2}{\sin(\theta_M)^2}\cdot\Big(\frac{\partial^2F}{\partial\theta_M^2}\frac{\partial^2F}{\partial\phi_M^2}-\big(\frac{\partial^2F}{\partial\theta_M\partial\phi_M}\big)^2\Big).
\label{eqFMR}
\ee 
Here $\theta_M$ is the angle of the magnetization with respect to the film's normal, $\phi_M$ is the in-plane angle of the magnetization with respect to the $[1\overline{1}0]$ crystallographic axis and $F=\frac{F'}{M_\text s}$, with $F'$ the free energy density and $M_\text s$ the saturation magnetization (Fig.~\ref{figaxis}). $\gamma=\frac{g\mu_\text B}{\hbar}$ is the gyromagnetic ratio, with $\mu_\text B$ the Bohr magneton and $\hbar$ the reduced Planck constant. The anisotropic g-factor is given by $g=\sqrt{g_\perp^2\cos(\theta_M)^2+g_{||}^2\sin(\theta_M)^2}$, with $g_{||}$ and $g_\perp$ respectively the in-plane and out-of-plane g-factors \cite{Farle1998}.\\      
For (111)-oriented films with cubic and uniaxial out-of-plane magnetic anisotropies the normalized free energy density is given by \cite{Manuilov2009,Dubs2020} 
\be
\begin{split}
&F=-B_0\Big(\sin(\theta_M)\sin(\theta_B)\cos(\theta_M-\theta_B)+\cos(\theta_M)\cos(\theta_B)\Big)+\frac{1}{2}\big(\mu_0M_\text s-\frac{2K_{2\perp}}{M_\text s}\big)\cos^2(\theta_M)\\
&+\frac{1}{2}\cdot\frac{2K_4}{M_\text s}\Big(\frac{1}{3}\cos^4(\theta_M)+\frac{1}{4}\sin^4(\theta_M)-\frac{\sqrt{2}}{3}\sin^3(\theta_M)\cos(\theta_M)\sin(3\phi_M)\Big),
\end{split}
\label{eqFree}
\ee  
with $\theta_B$ and $\phi_B$ the angles of $B_0$ with respect to respectively the film's normal and the in-plane $[1\overline{1}0]$ crystallographic axis (Fig.~\ref{figaxis}) and $\mu_0$ the vacuum permeability. $\frac{2K_{2\perp}}{M_\text s}$ and $\frac{2K_4}{M_\text s}$ are respectively the uniaxial out-of-plane and cubic anisotropy fields, with $K_{2\perp}$ and $K_4$ the perpendicular and cubic anisotropy constants. Note that to calculate the FMR frequency using equation~\ref{eqFMR} at a certain $B_0$, $\theta_B$ and $\phi_B$, we first need to find $\theta_M$ and $\phi_M$ that minimize the free energy by numerically solving $\frac{\partial F}{\partial\theta_M}(\theta_M,\phi_M)=0$ and $\frac{\partial F}{\partial\phi_M}(\theta_M,\phi_M)=0$.  \\
Using equations~\ref{eqFMR} and~\ref{eqFree} we can calculate the FMR frequency for an out-of-plane magnetic field and magnetization ($\theta_B=\theta_M=0^\circ$), which gives
\be
\omega_\text{FMR($\perp$)}=\gamma_{\perp}(B_0-\mu_0M_\text s+\frac{2K_{2\perp}}{M_\text s}-\frac{4K_{4}}{3M_\text s}).
\label{eqOOP}
\ee
For an in-plane magnetic field and magnetization ($\theta_B=\theta_M=90^\circ$), we find
\be
\omega_\text{FMR($||$)}=\gamma_{||}\sqrt{B_0\cdot\big(B_0+\mu_0M_\text s-\frac{2K_{2\perp}}{M_\text s}-\frac{K_{4}}{M_\text s}\big)-2\big(\frac{K_{4}}{M_\text s}\cos(3\phi_M)\big)^2}.
\label{eqipfmr}
\ee
The factor 3 in the cosine arises from the triangular in-plane symmetry of a cubic unit cell with its normal along the $[111]$ direction (Fig.~\ref{figaxis}). In our measurements a large in-plane magnetic field is needed to overcome the perpendicular anisotropy and push the magnetization in the plane, such that generally $B_0\gg|\frac{2K_{4}}{M_\text s}|=8.2$ mT and we can ignore the last term \cite{Manuilov2009}
\be
\omega_\text{FMR($||$)}=\gamma_{||}\sqrt{B_0\cdot(B_0+\mu_0M_\text s-\frac{2K_{2\perp}}{M_\text s}-\frac{K_{4}}{M_\text s})}.
\label{eqIP}
\ee
Equations~\ref{eqOOP} and~\ref{eqIP} are the same as equations~\ref{eq1} and~\ref{eq2} in the main text.
\begin{figure}[h!]
	\includegraphics[width=0.8\textwidth]{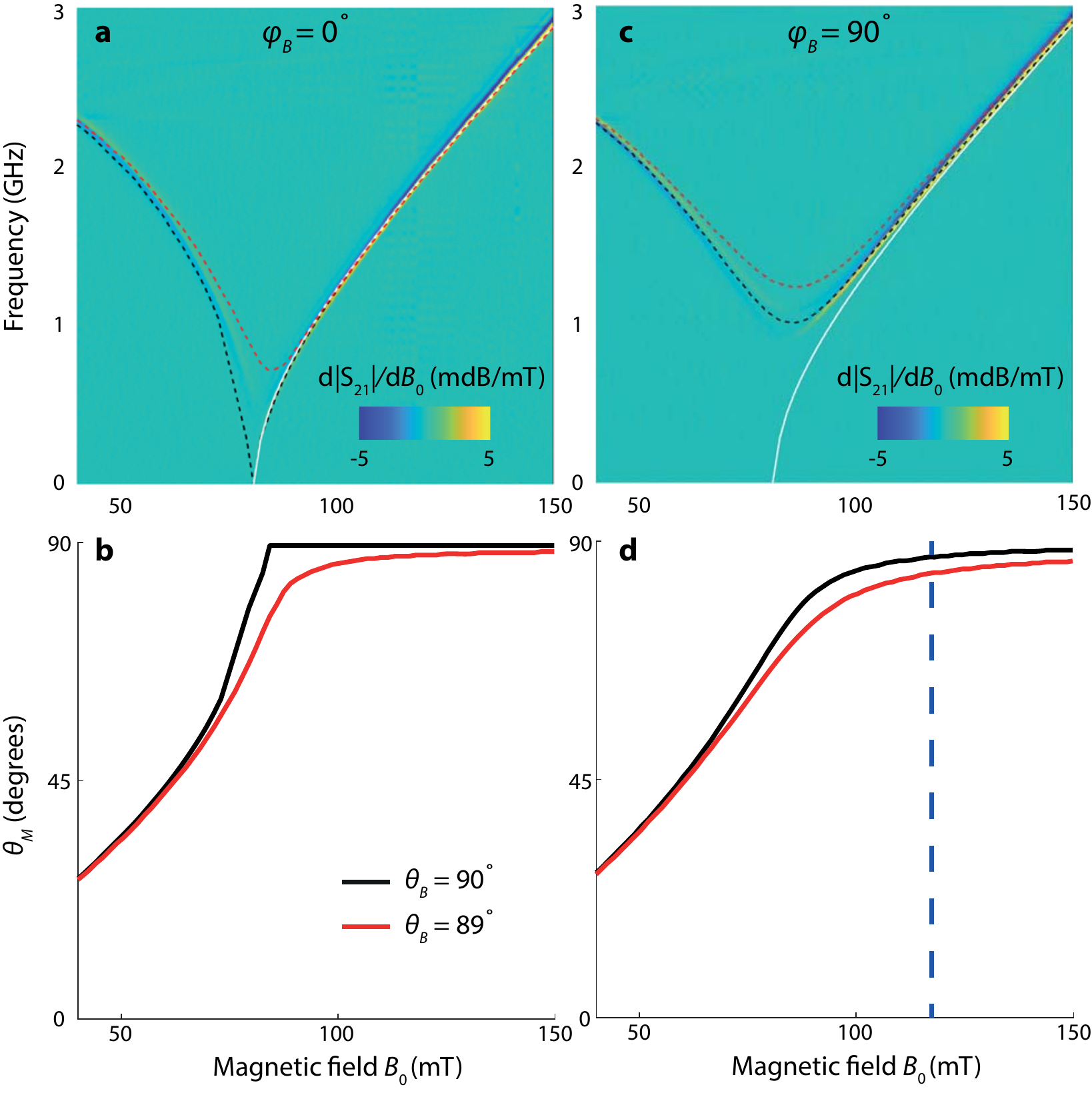}
	\caption{\textbf{Dependence of the FMR frequency on the direction of the external magnetic field $\bm{B_0}$.} (a) Flipchip FMR measurement with $B_0$ applied parallel to a 180-$\upmu$m-wide excitation stripline and along the in-plane $[1\overline{1}0]$ crystallographic direction ($\phi_B=0^\circ$). The FMR is extracted from the field-derivative of the microwave transmission $|\text{S}_{21}|$. The solid white line shows a fit to the Kittel relation (equation~\ref{eqIP}). Using the extracted anisotropy fields and gyromagnetic ratio, the FMR frequency for the entire $B_0$-range was numerically calculated assuming $\theta_B=90^\circ$ (black dashed line) and $\theta_B=89^\circ$ (red dashed line). (b) The minimum FMR frequency is raised at $\theta_B=89^\circ$ because the magnetization does not abruptly turn into the plane. (c) Similar FMR measurement, but with $B_0$ applied along the $[11\overline{2}]$ direction ($\phi_B=90^\circ$). The white line is the same as in (a). The black and red dashed lines are the calculated FMR frequencies for $\phi_B=90^\circ$ at respectively $\theta_B=90^\circ$ and $\theta_B=89^\circ$ using the parameters extracted in (a). (d) The magnetization maintains a finite out-of-plane component even when $\theta_B=90^\circ$. The blue dashed line indicates the field at which the exchange constant was determined from the spin-wave spectra. In (a) and (c) a similar background subtraction was performed as in Fig.~\ref{figBG}.              
	}
	\label{figFMR}
\end{figure}
\subsubsection{FMR frequency and magnetization direction at $\phi_B=0^\circ$}\label{Supp:0}
Fig.~\ref{figFMR}a shows a flipchip FMR measurement with the magnetic field applied along the $[1\overline{1}0]$ direction ($\theta_B=90^\circ$, $\phi_B=0^\circ$). The solid white line shows a fit to equation~\ref{eqIP} for magnetic fields at which the FMR frequency is increasing. Together with the fitted out-of-plane FMR, we extract $\frac{2K_{2\perp}}{M_\text s}=104.7$ mT, $\frac{2K_4}{M_\text s}=-8.2$ mT and $\frac{\gamma_{||}}{2\pi}=28.56$ MHz/mT. The same fit and data are presented in Fig.~\ref{fig1}b of the main text. \\
We can calculate the FMR frequency also for low bias fields by substituting the extracted parameters into equations~\ref{eqFMR} and~\ref{eqFree}. We obtain the black dashed line, which fits reasonably well to the measured FMR, even when the FMR frequency is decreasing with field. The red dashed line shows the calculated FMR frequency when $B_0$ has an $1^\circ$ out-of-plane angle ($\theta_B=89^\circ$), which dramatically increases the minimum FMR frequency. This is because the magnetization turns only asymptotically into the plane when the angle is offset, instead of abruptly (Fig.~\ref{figFMR}b, black line: $\theta_B=90^\circ$, red line: $\theta_B=89^\circ$). \\
We note that in Fig.~\ref{figFMR}a at large bias fields both the black and red dashed lines overlap with the white fit. Therefore, we conclude that the in-plane FMR at $\phi_B=0^\circ$ is quite robust to any small out-of-plane component of the static field that might be present in our experimental setup, validating the white fit using equation~\ref{eqIP} \cite{Manuilov2009}. 

\subsubsection{FMR frequency and magnetization direction at $\phi_B=90^\circ$}    
Fig.~\ref{figFMR}c shows a similar flipchip FMR measurement as in Fig.~\ref{figFMR}a, but now with the field applied along the $[11\overline{2}]$ direction ($\theta_B=90^\circ$, $\phi_B=90^\circ$, the white line is the same as in Fig.~\ref{figFMR}a and is added as a reference). The FMR reaches a minimum frequency of about 1 GHz, which is significantly larger than the minimum in the $\phi_B=0^\circ$ geometry. We reproduce this enhanced frequency minimum by calculating the expected FMR frequency using the parameters extracted in section~\ref{Supp:0} (black dashed line, we ignore any potential in-plane anisotropy of the g-factor). The calculated FMR frequency matches the measured FMR remarkably well for all magnetic field values, demonstrating the accuracy of the white fit. \\  
Again we attribute the enhanced FMR minumum to the fact that the magnetization only slowly turns into the plane, even for a perfect in-plane magnetic field $\theta_B=90^\circ$ (Fig.~\ref{figFMR}d, black line). As a result the FMR frequency asymptotically approaches the in-plane Kittel relation (equation~\ref{eqIP}, white line). Similar to before, a change of $1^\circ$ in $\theta_B$ lifts the minimum FMR frequency, explaining the minimum FMR frequency of about 1.25 GHz observed in Fig.~\ref{fig2}b in the main text. Variations on the order of $1^\circ$ in $\theta_B$ are expected in our measurement setup since we manually place the sample between two permanent magnets (section~\ref{Sup:GaYIG}).\\
Fig.~\ref{figFMR}d shows that the magnetization does not point exactly in the plane during our propagating spin wave spectroscopy measurements, even though this is assumed in the data analysis. We derived the exchange constant from spin-wave spectra taken at approximately $B_0=117.5$ mT, at which the magnetization points $\sim$3-6 degrees out of the plane (blue dashed line in Fig.~\ref{figFMR}d). We neglect this small out-of-plane angle, because we expect the induced error to be negligible compared to the $\sim15\%$ error obtained from the fit in Fig.~\ref{fig3}c in the main text.
\subsection{Systematic error in the applied bias field}\label{Sup:meas}
\begin{figure}[h!]\centering
	\includegraphics[]{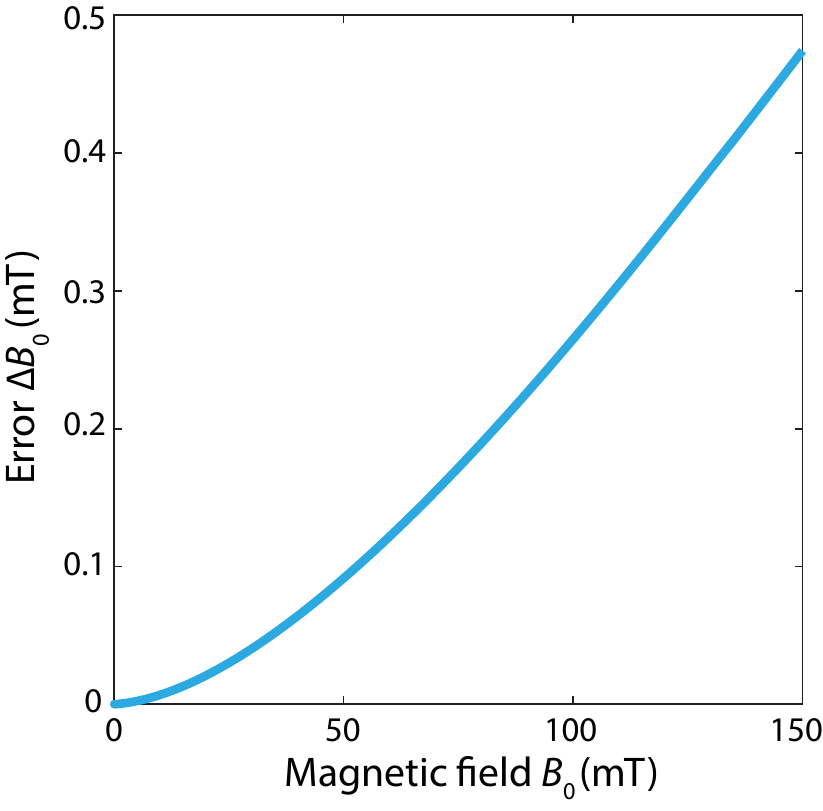}
	\caption{\textbf{Error in the static magnetic field as a result of a 1 mm offset of the sample with respect to the center position between the magnets.} For magnetic fields between 100 mT and 150 mT an error of $\sim$ 0.3-0.5 mT is expected.                
	}
	\label{figDB}
\end{figure}

In this section we calculate how a systematic error in the applied bias field affects the error of the anisotropy fields, which we extracted from the FMR frequency (Fig.~\ref{fig1}b of the main text). From the fits of the FMR frequency we obtain $\gamma_\perp=29.40(3)$ MHz/mT and $\alpha=-\mu_0M_\text s+\frac{2K_{2\perp}}{M_\text s}-\frac{2}{3}\frac{2K_4}{M_\text s}=91.1(2)$ mT (out-of-plane bias field), $\gamma_{||}=28.56(4)$ MHz/mT and $\beta=\mu_0M_\text s-\frac{2K_{2\perp}}{M_\text s}-\frac{1}{2}\frac{2K_4}{M_\text s}=-81.5(1)$ mT (in-plane bias field). Since we know from vibrating sample magnetometry (VSM) that $M_\text s=1.52(6)\cdot10^4$ A/m, we can calculate the magnetic anisotropy fields
\be
\begin{split}
&\frac{2K_4}{M_\text s}=-\frac{6}{7}(\alpha+\beta)=8.2(2) \text{ mT},\\
&\frac{2K_{2\perp}}{M_\text s}=\mu_0M_\text s+\frac{3}{7}\alpha-\frac{4}{7}\beta=104.7(8) \text{ mT}.
\end{split}
\ee
Since we manually place our sample between the magnets (section~\ref{Sup:GaYIG}), it may have a small offset of $\sim1$ mm with respect to the center position. Such an offset would cause a systematic error in the applied magnetic field $B_0$, which enhances the error of the anisotropy fields. To obtain a conservative estimate of these errors, we determine the systematic error in the applied magnetic field via 
\be
\Delta B_0(x)=B_0(x+1)+B_0(x-1)-2\cdot B_0(x).
\ee
$B_0(x)$ is the magnetic field of a cylindrical magnet at a distance of $x$ mm along its symmetry axis
\be
B_0(x)=\frac{B_r}{2}\Big(\frac{x+L}{\sqrt{r^2+(x+L)^2}}-\frac{x}{\sqrt{r^2+x^2}}\Big).
\ee 
Here $B_r=1320$ mT is the remanence, $L=20$ mm and $r=17.5$ mm are the length and radius of the magnet. Fig.~\ref{figDB} shows the calculated error $\Delta B_0(x)$ for a 1-mm-offset against the magnetic field $B_0$ at the center position between the magnets. Including this error in the fit of Fig.~\ref{fig1}b in the main text gives $\alpha=91.1(3)$ mT and $\beta=-81.5(5)$ mT, resulting in a slight increase in the error of the cubic anisotropy field $\frac{2K_4}{M_\text s}=8.2(5) \text{ mT}$. The errors in the gyromagnetic ratios and perpendicular anisotropy field do not change significantly. At the magnetic field $B_0=117.5$ mT at which we took the spin-wave spectra the error in the field $\Delta B_0$ is $\sim0.3$ mT.    
\subsection{Background-subtraction procedures of the spin-wave spectra}\label{Sup:BG}
\begin{figure}[h!]
	\includegraphics[width=1\textwidth]{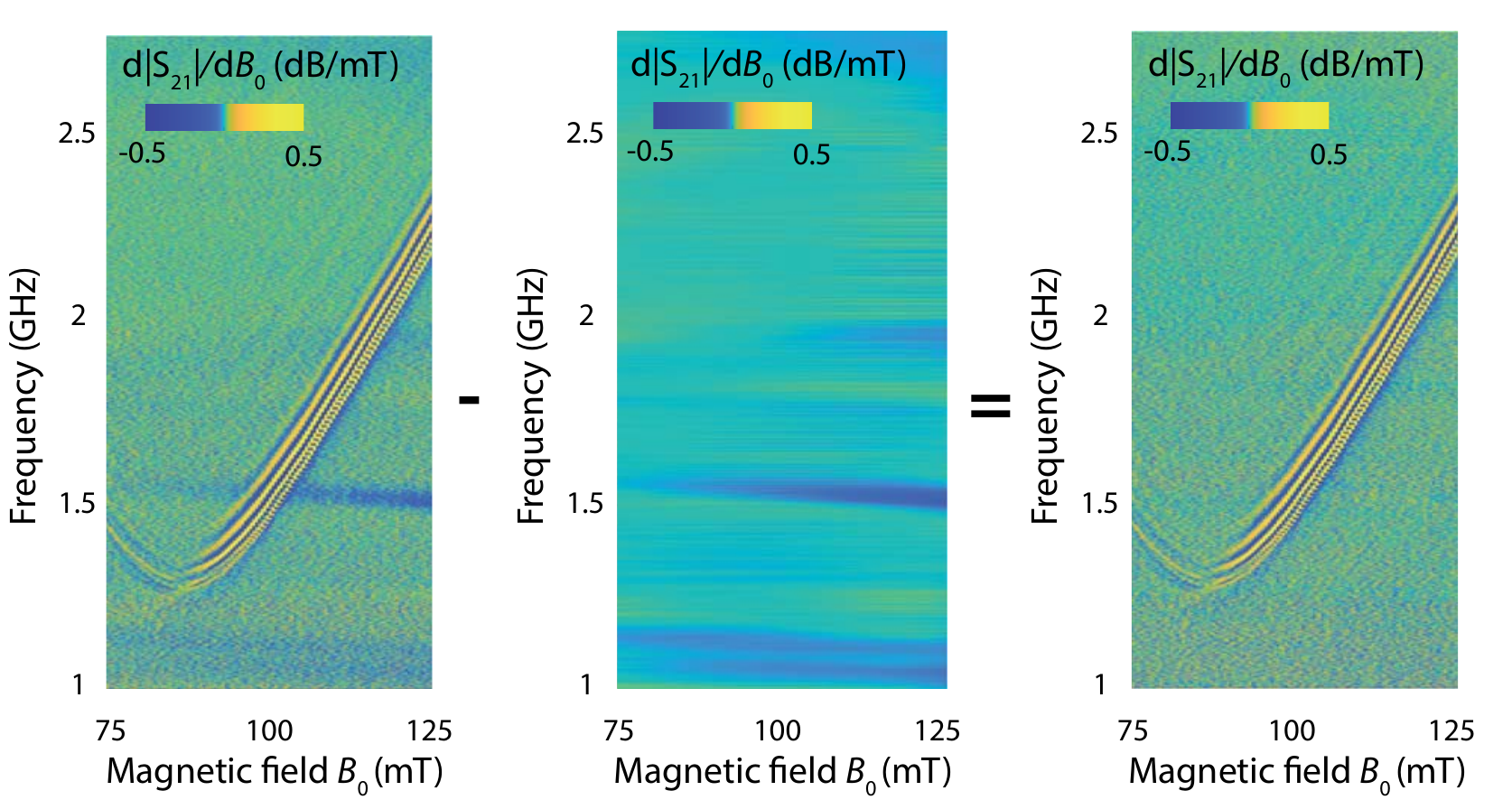}
	\caption{\textbf{Background-subtraction procedure of the microwave spectrum in Fig.~\ref{fig2}b of the main text.} The measured data (left figure) contains spurious signals attributed to small changes in the microwave transmission of the cables and connectors that attach the VNA to the striplines as a function of magnetic field. We filter these signals by first masking the high-curvature part of measured data that contains the spin-wave fringes. Then we fit a fifth-order polynomial through each horizontal line, excluding the masked data, and subtract it as a background (middle figure). The resulting spectrum only contains the spin-wave fringes (right figure, same as Fig.~\ref{fig2}b in the main text). The image processing was performed using Gwyddion (version 2.58).}
	\label{figBG}
\end{figure}

For the spin-wave spectra in Fig.~\ref{fig3}a,b and Fig.~\ref{fig4}a,c in the main text a background spectrum was subtracted consisting of the mean $|\text{S}_{21}|$ transmission at 100 mT and 138 mT, for which there are no spin waves in the frequency range of interest. In Fig.~\ref{fig2}b in the main text a background was subtracted using Gwyddion (Fig.~\ref{figBG}).    

\subsection{The spin-wave dispersion of a magnetic thin film with perpendicular and cubic magnetic anisotropy}\label{Sup:disp}

The spin-wave dispersion for magnetic thin films with perpendicular magnetic anisotropy (PMA) and cubic anisotropy was derived in reference \cite{Kalinikos1990}. Equation 30 of this work states the dispersion for an (111)-oriented film with in-plane magnetization, similar as in our experiment
\be
\begin{split}
&\omega_\text{SW}(k)=\\
&\gamma_{||}\sqrt{\big(B_0+Dk^2+\mu_0M_\text s(1-f)-\frac{2K_{2\perp}}{M_\text s}-\frac{K_{4}}{M_\text s}\big)\cdot\big(B_0+Dk^2+\mu_0M_\text sf\sin^2(\phi)\big)-2\big(\frac{K_{4}}{M_\text s}\cos(3\phi_M)\big)^2}.
\end{split}
\label{eqdisp}
\ee
Here $\omega_\text{SW}$ is the angular frequency of a spin wave with wavevector $k$ that propagates at an angle $\phi$ with respect to the magnetization. $D=2\alpha/M_\text s$ is the spin stiffness, with $\alpha$ the exchange constant, and $f=1-(1-e^{-kt})/kt$ with $t$ the thickness of the film and $\phi_M$ is the angle of the magnetization with respect to $[1\overline{1}0]$ crystallographic direction. We note that if we set $k=0$ in equation~\ref{eqdisp}, we obtain the in-plane FMR frequency derived before (equation~\ref{eqipfmr}). \\
In our experiment we measure spin waves in the Damon-Eshbach configuration ($\phi=\pi/2$), we apply the external field $B_0$ along $[11\overline{2}]$ ($\phi_M=\pi/2$) and the wavelengths of the detected spin waves are much smaller than the thickness of the film ($kt\ll1$), such that we can approximate $f\approx kt/2$. This gives
\be
\omega_\text{SW}(k)=\sqrt{\big(\omega_B+\gamma_{||}Dk^2-\omega_K+\omega_M(1- kt/2)\big)\big(\omega_B+\gamma_{||}Dk^2+\omega_M  kt/2\big)}, 
\ee
where we defined $ \omega_B = \gamma_{||} B_\text{0}$, $\omega_M =  \gamma_{||} \mu_0 M_\text s$, and $\omega_K =\gamma_{||}( \frac{2K_{2\perp}}{M_\text s}+\frac{K_{4}}{M_\text s})$ for convenience of notation. Working out the brackets and rearranging the terms in orders of $k$ gives
\be
\omega_\text{SW}=\sqrt{\omega_B\big(\omega_B+\omega_M-\omega_K\big)+\frac{\omega_Mt}{2}\big(\omega_M-\omega_K\big)k+\gamma_{||}D\big(2\omega_B+\omega_M-\omega_K-(\frac{\omega_Mt}{2})^2\big)k^2+\gamma_{||}^2D^2k^4}.
\ee
For the spin-wave spectra taken at $B_0=117.5$ mT we find  ($\frac{\omega_Mt}{2})^2\ll2\omega_B+\omega_M-\omega_K$ due to the low saturation magnetisation and thickness of our film, such that we can further approximate
\be
\omega_\text{SW}=\sqrt{\omega_B(\omega_B+\omega_M-\omega_K)+\frac{\omega_Mt}{2}(\omega_M-\omega_K)k+\gamma_{||}D(2\omega_B+\omega_M-\omega_K)k^2+\gamma_{||}^2D^2k^4},
\ee
 which is equation~\ref{eq4} in the main text. \\
We derive the group velocity $v_\text g$ by differentiating with respect to $k$
\be
v_\text g=\frac{\partial\omega_\text{SW}}{\partial k}=\frac{1}{2\sqrt{\omega_\text{SW}}}\big(\frac{\omega_Mt}{2}(\omega_M-\omega_K)+2\gamma_{||}D(2\omega_B+\omega_M-\omega_K)k+4\gamma_{||}^2D^2k^3\big),
\ee
 which is equation~\ref{eq5} in the main text.

\subsection{Comparing the frequency difference between fringes to the spin-wave linewidth}\label{Sup:line}
In this section we calculate the expected average frequency difference $\Delta f$ between spin-wave fringes excited by the first maximum of the microwave driving field Fourier amplitude ($|B^{\text{RF}}(k)|$) in Fig.~\ref{fig3}b of the main text. The stripline has a width $w=2.5$ $\upmu$m, such that $|B^{\text{RF}}(k)|$ has its first node at $k_{\text{min}}=\frac{2\pi}{2.5}$ $\upmu\text{m}^ {-1}$ \cite{Ciubotaru2016}. Everytime another wavelength fits within the center-to-center distance $s$ between both striplines another fringe is observed in the signal. Therefore the condition $s=n\lambda$ applies for every $n$th fringe, with $\lambda$ the spin-wave wavelength. This means that fringes occur every $\Delta k=\frac{2\pi}{s}= \frac{2\pi}{12.5}$ $\upmu\text{m}^ {-1}$ in k-space. In the first maximum of the excitation spectrum we would thus expect $\frac{k_{\text{min}}}{\Delta k}=5$ fringes. According to the reconstructed dispersion (Fig.~\ref{fig3}f of the main text) the frequency difference between spin waves with wavevector $k_{\text{min}}$ and the minimum of the band is about $20$ MHz, leading to an average frequency difference of $\frac{20}{5}=4$ MHz between consecutive fringes. This is on the order of the FMR linewidth of undoped YIG films of similar thicknesses \cite{Dubs2020}. Assuming that Ga:YIG has a similar or larger linewidth, we argue that we cannot resolve fringes in the first maximum of the excitation field's Fourier amplitude because they are too narrow compared to the intrinsic spin-wave linewidth.  
\subsection{Zoomed-in spin-wave spectra displaying low-amplitude fringes}\label{Sup:fring}  

\begin{figure}[h!]\centering
	\includegraphics[]{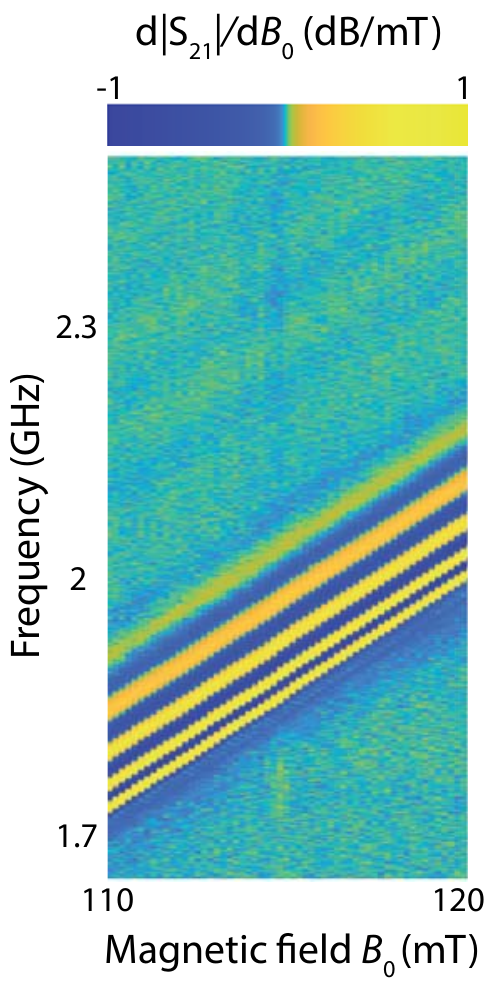}
	\caption{\textbf{Detailed microwave spectrum zoomed-in on the spin-wave fringes.} Low-amplitude fringes excited by the second maximum of the excitation field's Fourier amplitude are visible at high frequencies. The actual measured data without any background-subtraction is presented ($w=1$ $\upmu$m, $s=6$ $\upmu$m, excitation power -35 dBm).     
	}
	\label{fig}
\end{figure}

\subsection{Calculation of the non-linear frequency-shift coefficient}\label{Sup:freq}
For Damon-Eshbach spin waves with wavevector $k$ and frequency $\omega_k/2\pi$ the non-linear four-magnon frequency-shift coefficient $W_{kk,kk}$ is given by \cite{Krivosik2010}
\be
\begin{split}
&W_{kk,kk}=\frac{1}{2}\big(\frac{2\omega_B+\omega_M(N_{xx,k}+N_{yy,k})}{2\omega_k}\big)^2\cdot\big(3\omega_B+\omega_M(2N_{zz,0}+N_{zz,2k})\big)\\
&-\frac{1}{2}\big(3\omega_B+\omega_M(N_{xx,k}+N_{yy,k}+N_{zz,2k})\big),
\end{split}
\label{eqW}
\ee
with $N_{ij,k}$ the $(i,j)$th index of the spin-wave tensor $N_{k}$. The three-wave correction term vanishes since the spin waves propagate perpendicular to the magnetization. The precessional $xyz$-frame is defined such that $z$ points in the plane along the magnetization, $x$ along the film normal and $y$ points in-plane perpendicular to $z$ and parallel to the wavevector of the spin waves. \\
$N_k$ is the Fourier transform of the tensorial Green's function $N(r,r')=N(r,r')_\text{dip}+N(r,r')_\text{ex}+N(r,r')_\text{ani}$, which has components due to uniaxial anisotropy and the dipolar and exchange interactions
\be
N_ke^{ikr}=\int N(r,r')e^{ikr'}d^3r'=\int \big(N(r,r')_\text{dip}+N(r,r')_\text{ex}\big)e^{ikr'}d^3r'+\int N(r,r')_\text{ani}e^{ikr'}d^3r'.
\ee 
\\
The contribution to $N_k$ from the $N(r,r')_\text{dip}$ and $N(r,r')_\text{ex}$ components in the thin-film limit were derived earlier \cite{Krivosik2010}. Following this work, $ N(r,r')_\text{ani}$ due to uniaxial anisotropy in the out-of-plane $x$-direction is given by
\be
N(r,r')_\text{ani}=-\frac{B_{2\perp}}{\mu_0M_\text s}\delta(r-r')\hat{x}\otimes \hat{x}.
\ee    
Here $B_{2\perp}=\frac{2K_{2\perp}}{M_\text s}$ is the uniaxial out-of-plane anisotropy field, $\otimes$ denotes a dyadic unit vector product and $\delta(r-r')$ is the Dirac delta function. As a result of the dyadic product only the $(x,x)$ index of $ N(r,r')_\text{ani}$ is non-zero, leading to a contribution on $N_{xx,k}$
\be
N_{xx,k}e^{ikr}=\int -\frac{B_{2\perp}}{\mu_0M_\text s}\delta(r-r')e^{ikr'}d^3r'=-\frac{B_{2\perp}}{\mu_0M_\text s}e^{ikr}.
\ee
By adding this contribution to the other components, we find that the diagonal elements of $N_k$ in the Damon-Eshbach configuration are given by
\be
\begin{split}
&N_{xx,k}=\frac{D}{\mu_0 M_\text s}k^2+1-f-\frac{B_{2\perp}}{\mu_0M_\text s},\\
&N_{yy,k}=\frac{D}{\mu_0 M_\text s}k^2+f,\\
&N_{xx,k}=\frac{D}{\mu_0 M_\text s}k^2,\\
\end{split}
\label{eqN}
\ee 
with $f=1-(1-e^{-kt})/kt$ and $t$ the thickness of the film as before. We neglected the cubic anisotropy since it is small relative to the uniaxial anisotropy.\\
\begin{figure}[h!]\centering
	\includegraphics[]{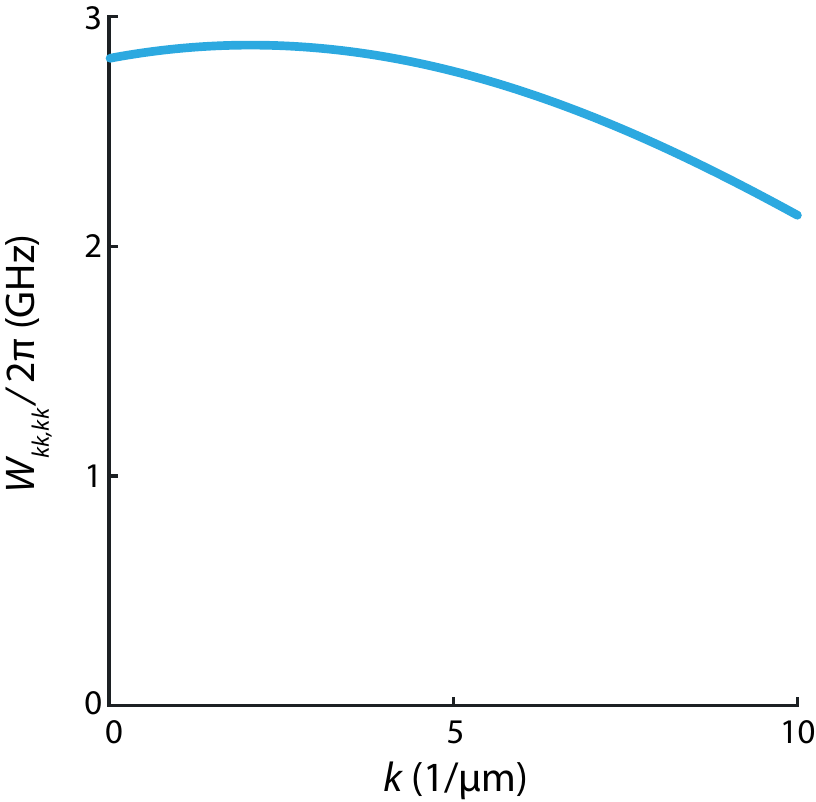}
	\caption{\textbf{Non-linear frequency-shift coefficient $\bm{W_{kk,kk}}$ for Damon-Eshbach spin waves in Ga:YIG.} We used the dispersion in Fig.~\ref{fig3}f of the main text as an input, together with the extracted parameters $B_0=117.5$ mT, $\alpha=1.3\cdot10^{-12}$ J/m, $\frac{2K_{2\perp}}{M_\text s}=104.7$ mT, $\frac{\gamma_{||}}{2\pi}=28.56$ MHz/mT, $M_\text s=1.52\cdot10^{-4}$ A/m and $t=45$ nm. The cubic anisotropy is neglected. The positive sign of the calculated frequency-shift coefficient matches the positive frequency shifts observed in the experiment.              
	}
	\label{figW}
\end{figure}
\\
By substituting equations~\ref{eqN} into equation~\ref{eqW} we can calculate $W_{kk,kk}$ for the wavevectors relevant for this work (Fig.~\ref{figW}). For all these wavevectors $W_{kk,kk}$ is positive, explaining the positive frequency shifts of the spin waves that we observe when increasing the drive power. This is in contrast to the frequency shift caused by the reduction of the saturation magnetization as a result of strong driving or heating. In this simple picture a downward frequency shift is expected for in-plane magnetization (Fig.~\ref{figMs}), highlighting the value of the Hamiltonian formalism that was used to calculate the non-linear frequency-shift coefficient \cite{Krivosik2010}.\\
\\
\begin{figure}[h!]\centering
	\includegraphics[]{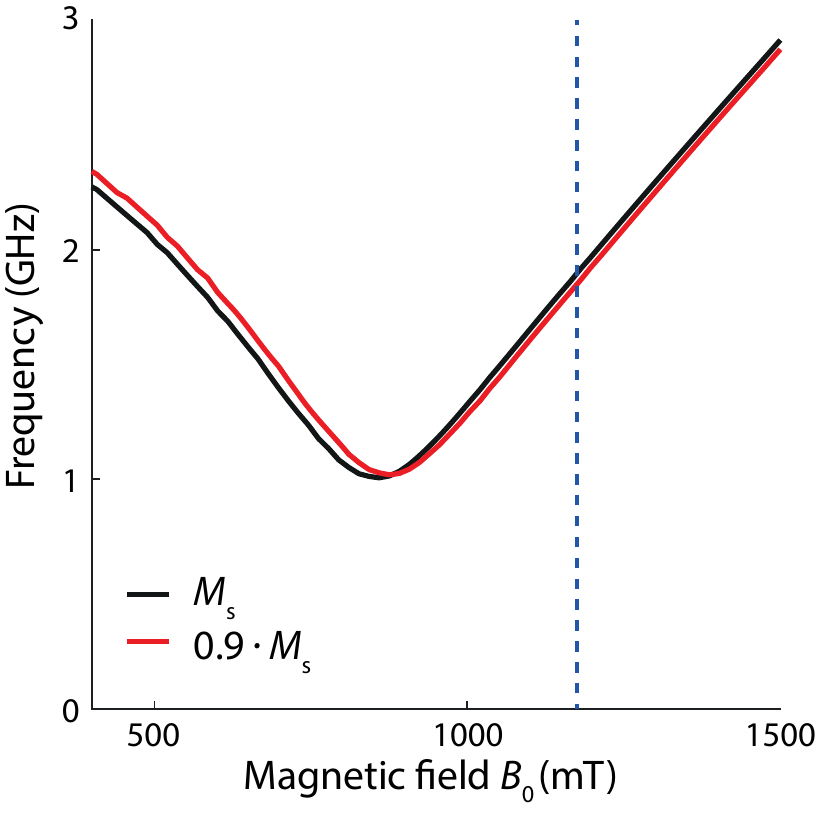}
	\caption{\textbf{Expected downward frequency shift upon reduction of the saturation magnetization.} Field dependence of the FMR frequency of Ga:YIG for unreduced saturation magnetization ($M_\text s=1.52\cdot10^4$ A/m, black line) and for 10\%-reduced saturation magnetization ($M_\text s=1.37\cdot10^4$ A/m, red line). The bias field is applied in the $[11\overline{2}]$ direction and the magnetic anisotropy fields are the same for both curves. The dashed line indicates the field at which we performed our spin-wave spectroscopy measurements. Clearly a negative frequency shift is expected upon decreasing the saturation magnetization, which is in contrast to the positive frequency shifts we observe.                 
	}
	\label{figMs}
\end{figure}
\printbibliography
\end{refsection}
\end{document}